\documentclass[12pt]{article}

\usepackage{graphicx} 

\begin{document}

\title{Z production via Vector Boson Fusion at LHC}
\vspace{1.5cm}
\author{P. Govoni$^a$ , C. Mariotti$^b$\\
{\small \sl $^a$ University of Milano Bicocca, $^b$ INFN Torino}}

\date{}
\maketitle

\begin{abstract}

The production of Z bosons via Vector Boson Fusion at the LHC collider 
at 10~TeV centre-of-mass energy has been studied. The aim is to
investigate the possibility to isolate a known Standard Model 
process to be used as reference for the measurement of the detector performance 
for the  search of the Higgs Boson produced via Vector Boson
Fusion.
The signal to background ratio has been estimated considering only
the dominant sources of background.

\end{abstract}

\section{Introduction} 

\begin{figure}[htbp]
 \centering
    \includegraphics[width = 0.3\textwidth]{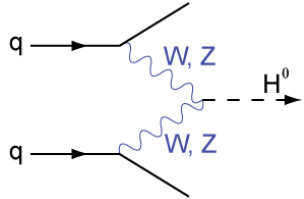} 
  \caption{The diagram of the VBF production of the SM
  Higgs boson at tree level.}
   \label{fig:VBF}
\end{figure}

The search of the Higgs boson is one of the principal goals of the experiments at the LHC. 
Depending on its mass, 
it will be searched for in various final states and with different production mechanisms. 
The production mechanism with a cross section only second to the gluon-gluon fusion one 
is the boson-boson fusion (VBF). 
This mechanism, as shown in fig. \ref{fig:VBF},  
allows a very clean signature of the final state. 
The incoming fermions scatter at high energy 
and enter the active volume of the LHC detectors in the high rapidity regions.
The presence of the forward and backward jets induced by the two fermions (here called ``tag jets'')  
characterizes the event as VBF 
(for a detailed study of the characteristic of these jets see for
example \cite{balle, ATLASTDR, CMSTDR}. 
The Vector Boson Fusion (VBF) production presents another important feature:
being a pure electroweak process 
the color exchange between the W and Z is minimal, 
hence the jet activity in the central part of the detector is very low, if not absent. 
These features are exploited to reject the backgrounds 
and identify the VBF sample among the data produced at hadron colliders, 
by requiring the events to have a pair of jets, 
in opposite directions with respect to the beam axis, in the forward-backward region, 
and no other jets in the central part of the detector.
To correctly establish the Higgs signal is therefore mandatory 
to know with good precision the identification efficiency of the forward-backward jets 
and the efficiency of the central jet veto.
The best way to determine these numbers from the data, without relying on the Monte Carlo simulation, 
would be to use a well know Standard Model process with similar kinematics.
The production of a single Z boson via VBF 
could be a powerful instrument to evaluate the sensitivity of LHC detectors 
to the VBF production of the Higgs boson \cite{dgreen}.

In this paper, a parton level feasibility study of the VBF Z production measurement at LHC is presented. 
After the signal definition, 
the cuts used to enhance the signal with respect to irreducible pure electroweak backgrounds are discussed;
finally, the dominant background for this analysis, 
coming from the production of a Z from $q \bar q$ fusion accompanied by two jets, 
is considered.
The unfortunately negative conclusion ends this work.

\section{The signal definition}

The signal events are Z bosons, that subsequently decay into leptons,
 produced via VBF.  
The generation of these events is done with the Matrix
Element Montecarlo Madgraph \cite{mad} 
asking for a final state with two leptons and two quarks.  
The process is at the order $\alpha_{EW}^4$.

\begin{figure}[htbp]
   \centering
   \includegraphics[width = 0.3\textwidth]{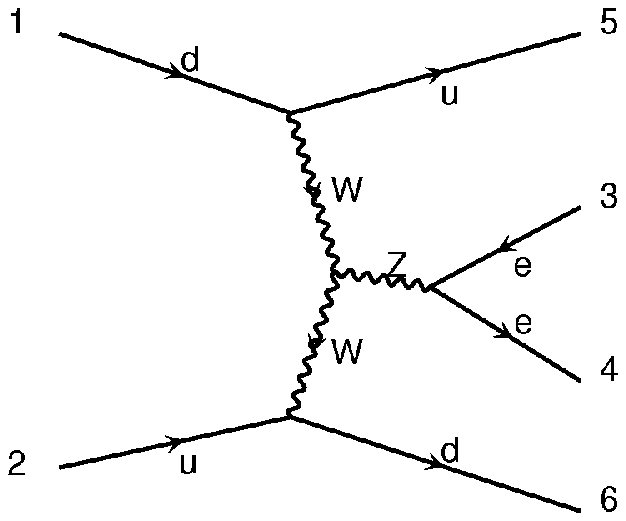} %
   \includegraphics[width = 0.3\textwidth]{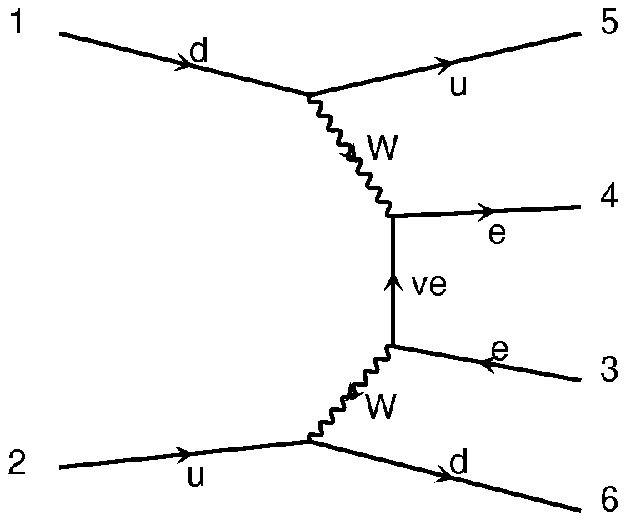} 
   \includegraphics[width = 0.3\textwidth]{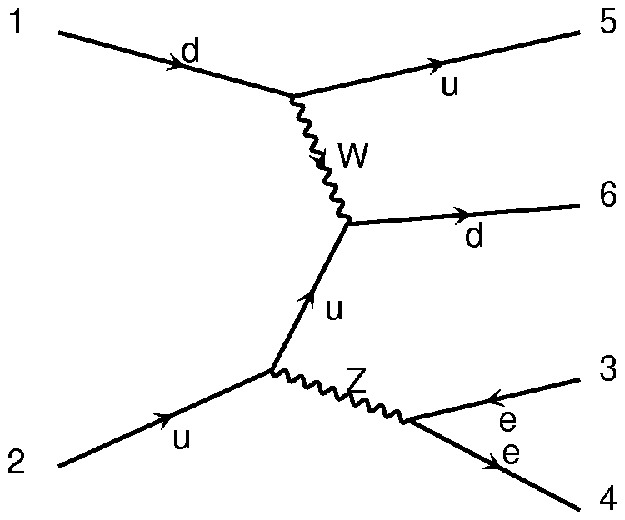} 
   \includegraphics[width = 0.3\textwidth]{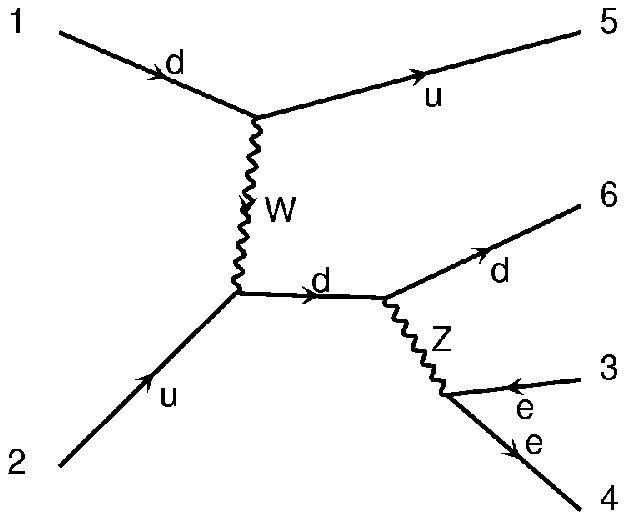} 
   \includegraphics[width = 0.3\textwidth]{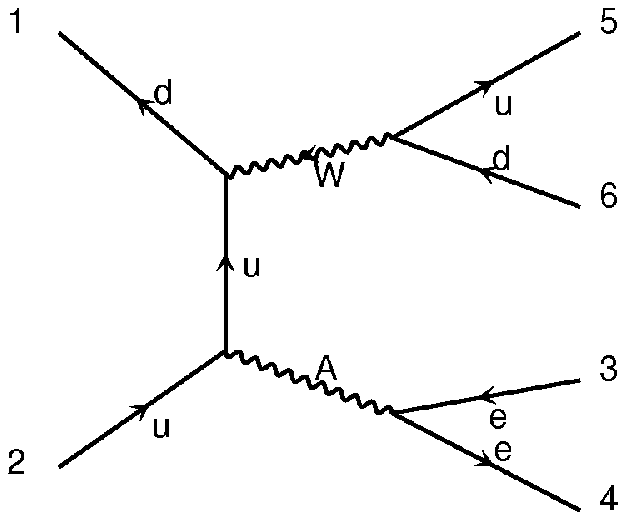} 
   \includegraphics[width = 0.3\textwidth]{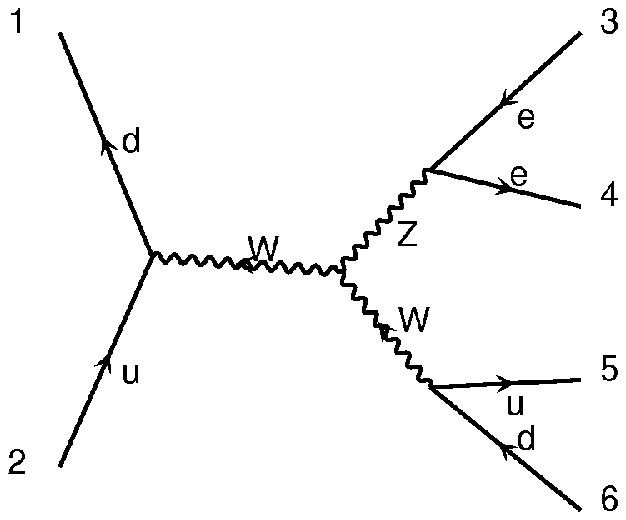} 
   \caption{A small subset of the graphs contributing to the four fermions final state
                  at the $\alpha_{EW}^4$ leading order.
                  The first diagram corresponds to the signal isolated with the selection
                  cuts listed in table \ref{tab:VBFMCcuts}.}
   \label{fig:graphSignal}
\end{figure}

A small subset of the diagrams that contribute to the production of
the four fermions final state at the $\alpha_{EW}^4$ order  ($ pp \rightarrow q_1 q_2 l^+l^-$)  
are shown in  figure \ref{fig:graphSignal}.
The total cross section is 
$\sigma =4024 \pm 2$~fb.

Out of the complete set of diagrams,
the first one in the figure corresponds to the VBF signal, which is searched for. 
The other diagrams will be hereafter called the electroweak irreducible background. 
For the purpose of this study, 
a set of selection cuts has been implemented at Montecarlo level
to isolate the signal from the other electroweak processes,
based on the invariant masses of the final state leptons 
and on the flavour and charge of the initial and final state quarks:
events surviving the selections listed in table \ref{tab:VBFMCcuts} constitute the signal sample,
while the others compose the irreducible electroweak background.
Final states with electrons or muons only have been considered.

The fraction of VBF like events is about 4.5\% of the total production (0.18~pb), 
that corresponds to about 36 events produced at 10~TeV with an integrated luminosity of 200~pb$^{-1}$.

\begin{table}[htbp]
   \centering
   \begin{tabular}{|c|} 
\hline
leptons ($\ell = e, \mu$) belong to the same family \\
\hline
86 GeV $<M_{\ell{}\ell{}}<$ 96 GeV \\
\hline
$M_{q_{out}q'_{out}}<$ 75 GeV or $M_{q_{out}q'_{out}}>$ 96 GeV \\
\hline
Incoming and outgoing quarks \\
should be compatible with emitted Ws\\
of opposite sign\\
\hline
   \end{tabular}
   \caption{MC-level selections to isolate the VBF contribution
                  from the other four fermions final state processes
                  at the $\alpha_{EW}^4$ leading order.}
   \label{tab:VBFMCcuts}
\end{table}

\begin{figure}[htbp]
   \centering
  \includegraphics[width = 0.55\textwidth]{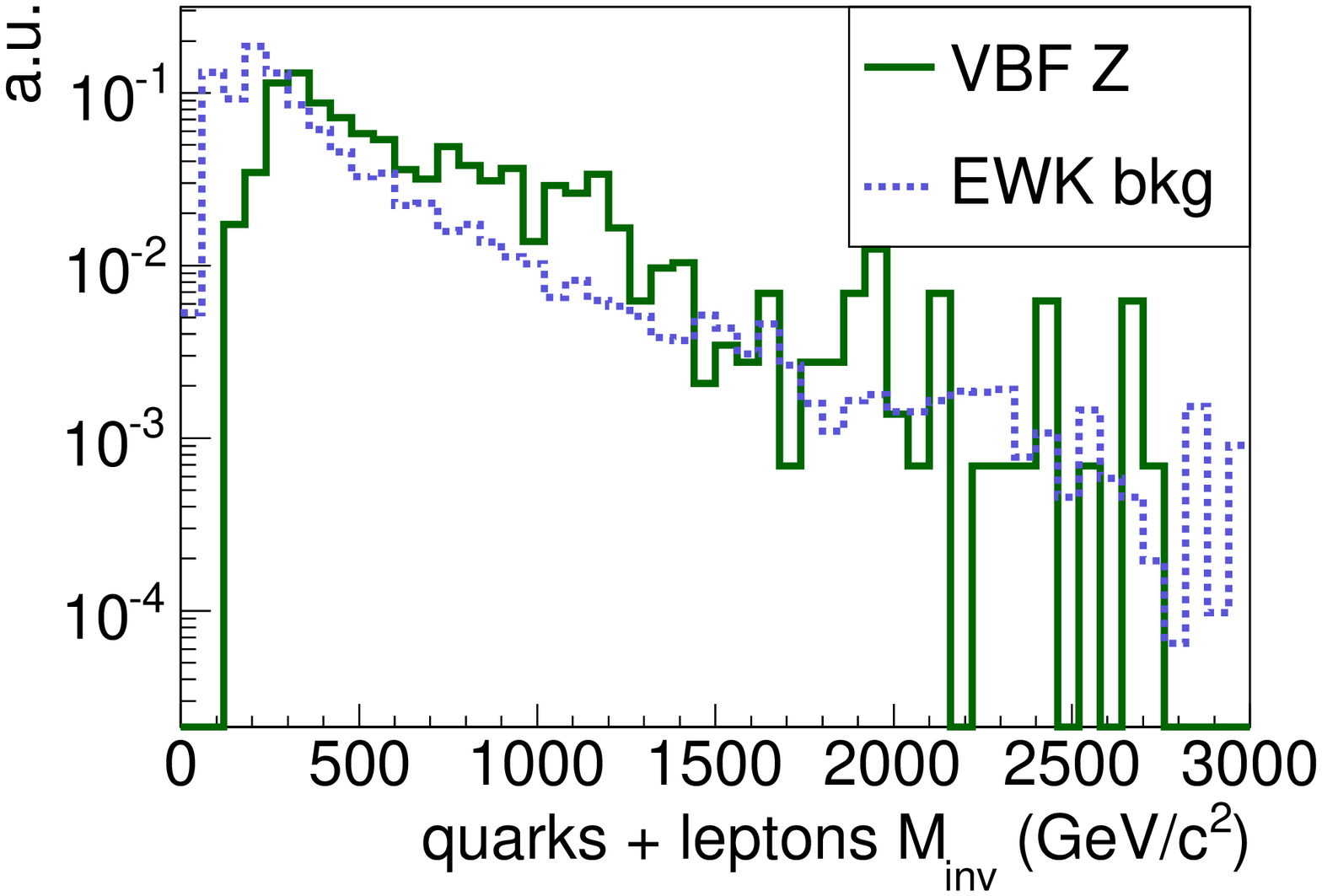} 
   \includegraphics[width = 0.55\textwidth]{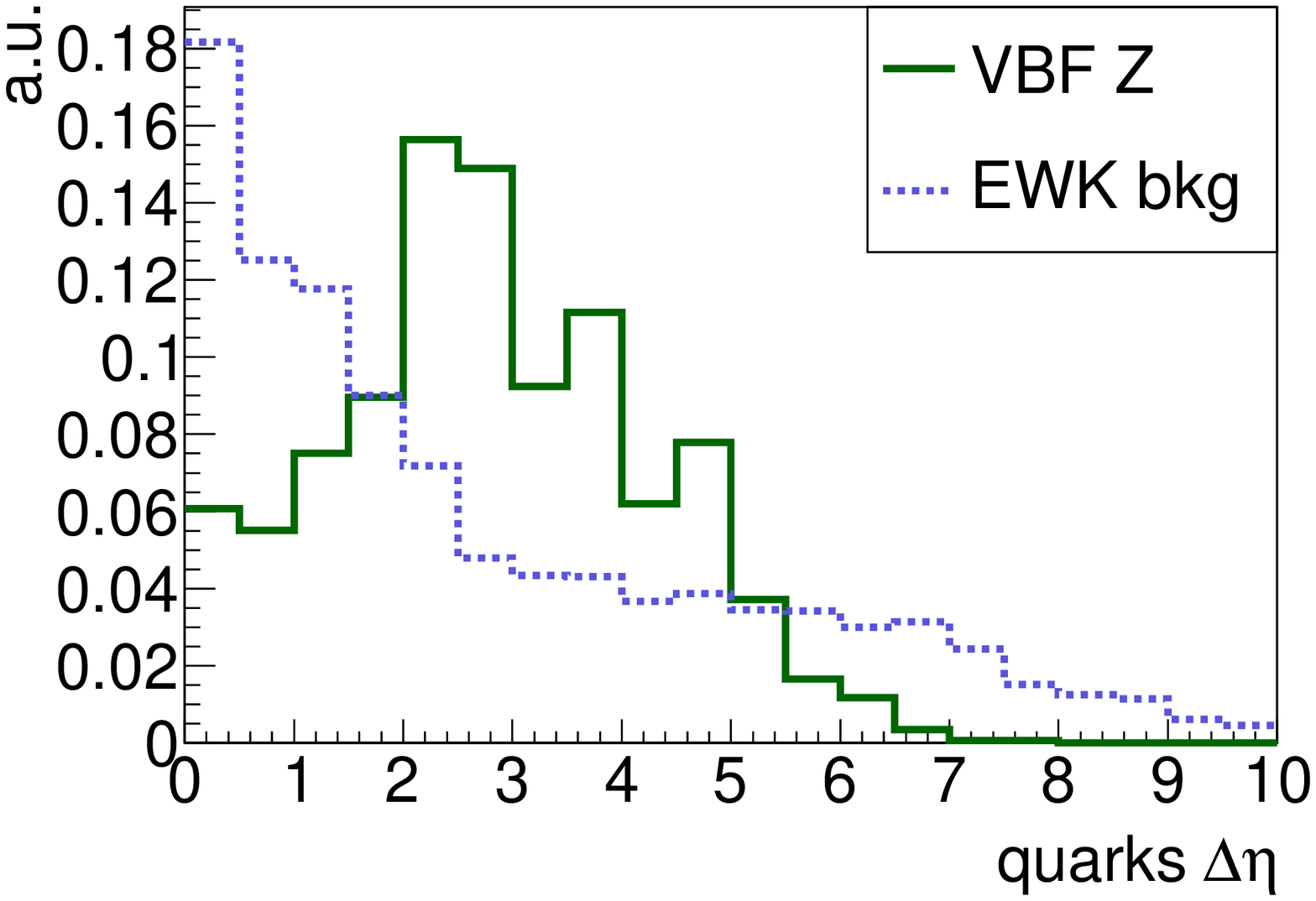} 
   \includegraphics[width = 0.55\textwidth]{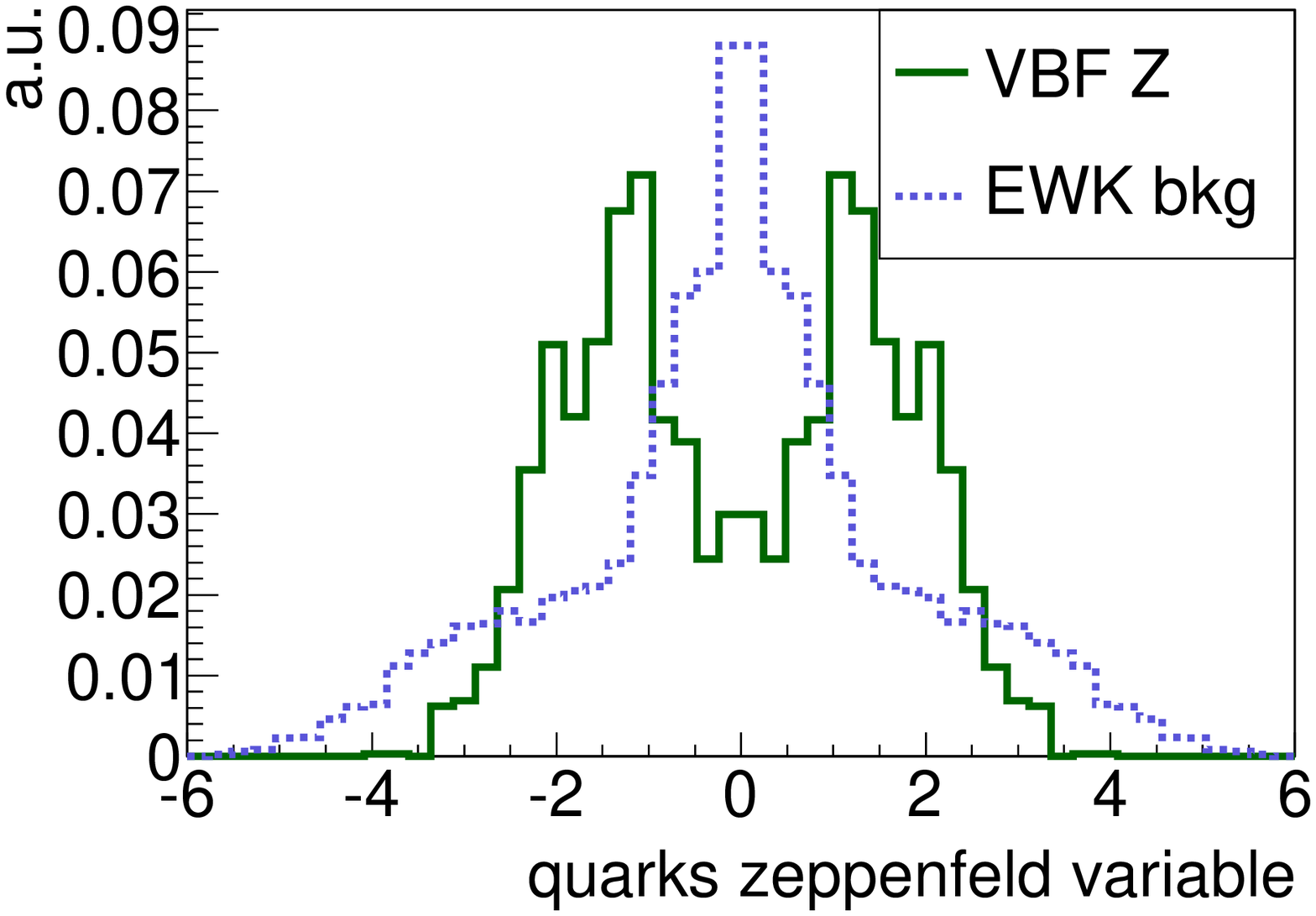} 
   \caption{The invariant mass of the four fermions in the final state, the
     $\Delta\eta_{jj}$ of the outgoing quarks and  Zeppenfeld variable
     are shown for the VBF signal (continuous green)  and the pure electroweak 
     backgrounds  (dashed blue) as defined by the selections listed in table \ref{tab:VBFMCcuts}.}
   \label{fig:lljjQED}
\end{figure}

The distributions of the invariant mass of the four final state fermions,
of the $\Delta\eta_{jj}$ between the tag jets,
and of their Zeppenfeld variable \cite{zeppenfeld}: 
\begin{equation}
\eta_i^*~=~\eta_i-\langle\eta\rangle_{1,2}
\end{equation}
are shown in figure \ref{fig:lljjQED}, for the
selected process and for the pure electroweak backgrounds.

\begin{figure}[htbp]
   \centering
   \includegraphics[width = 0.65\textwidth]{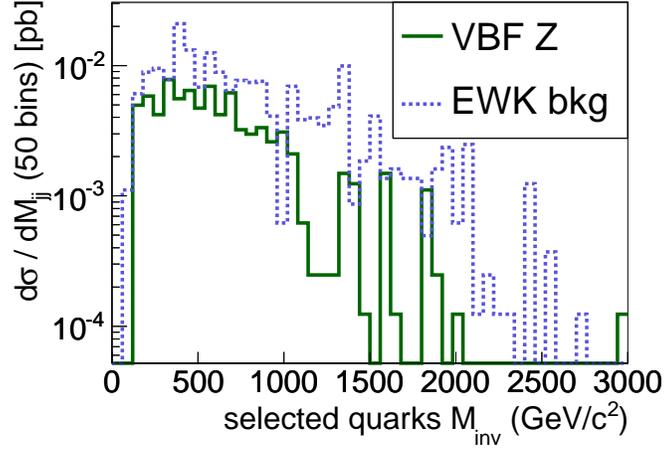} 
   \caption{The invariant mass of the two tag jets for the signal (full green) 
   and for the electroweak background (dashed blue), 
   after the selections listed in table \ref{tab:VBFRECOcuts}.}
   \label{fig:afterEWK_Mjj}
\end{figure}

As can be seen from the distributions,  the VBF signal contribution 
can be enhanced with respect to the irreducible background
thanks to the typical event topology.
Selection cuts based on the $\Delta\eta$ between the two jets,
on their transverse momentum, 
on their $\eta^*$, 
as well as the request on the invariant mass of the two leptons and two jets,
have been used, as summarized in 
table \ref{tab:VBFRECOcuts}.
\begin{table}[htbp]
   \centering
   \begin{tabular}{|c|c|c|} 
\hline
selection var & min & max \\
\hline
$|\Delta\eta_{jj}|$ & 1.5       & -- \\ 
min jet $p_T$       & 30 GeV    & -- \\
max jet $p_T$       & 50 GeV    & -- \\
$\eta_j^*$           & 1         & 3 \\
$M_{ll}$            & 86 GeV    & 96 GeV \\
$M_{jj}$            & 0 GeV     & 60 GeV \\
$M_{jj}$            & 111 GeV   & -- \\
$p_T(Z)$            & 30 GeV    &  -- \\
\hline
   \end{tabular}
   \caption{ Selection cuts to enhance the VBF signal contribution.}
   \label{tab:VBFRECOcuts}
\end{table}
With these selection cuts, 
an efficiency of 45\% for the signal, 
of 4.8\% for the electroweak background 
and a signal purity of 30\% can be reached for the
 $\alpha_{EW}^4$  order processes. 
 In figure \ref{fig:afterEWK_Mjj} the invariant mass of the two tag jets
is shown for the signal and the pure electroweak backgrounds after the selection cuts.
The effective cross sections of the processes after the selection cuts 
are summarized in table \ref{tab:VBFRECOcutsResults}.
\begin{table}[htbp]
   \centering
   \begin{tabular}{|l|c|c|c|} 
\hline
             & x-section & effective x-section & efficiency \\
             &           & after selections    &  \\
\hline
VBF signal   & 0.18 pb   & 0.082 pb            & 0.45 \\
EWK bkg      & 3.85 pb   & 0.19 pb             & 0.048 \\
\hline
S/B          &  0.05     &  0.43               & -- \\
\hline
   \end{tabular}
   \caption{VBF signal and EWK background cross sections, 
   before and after the selection cuts listed in table \ref{tab:VBFRECOcuts} 
   aiming to reduce the electroweak backgrounds.
   The signal and background are defined via the MC selections listed in table \ref{tab:VBFMCcuts}.
   The selection efficiencies and the signal to background (S/B) ratios are listed as well.}
   \label{tab:VBFRECOcutsResults}
\end{table}

\section{The dominant background}

\begin{figure}[htbp]
   \centering
   \includegraphics[width = 0.3\textwidth]{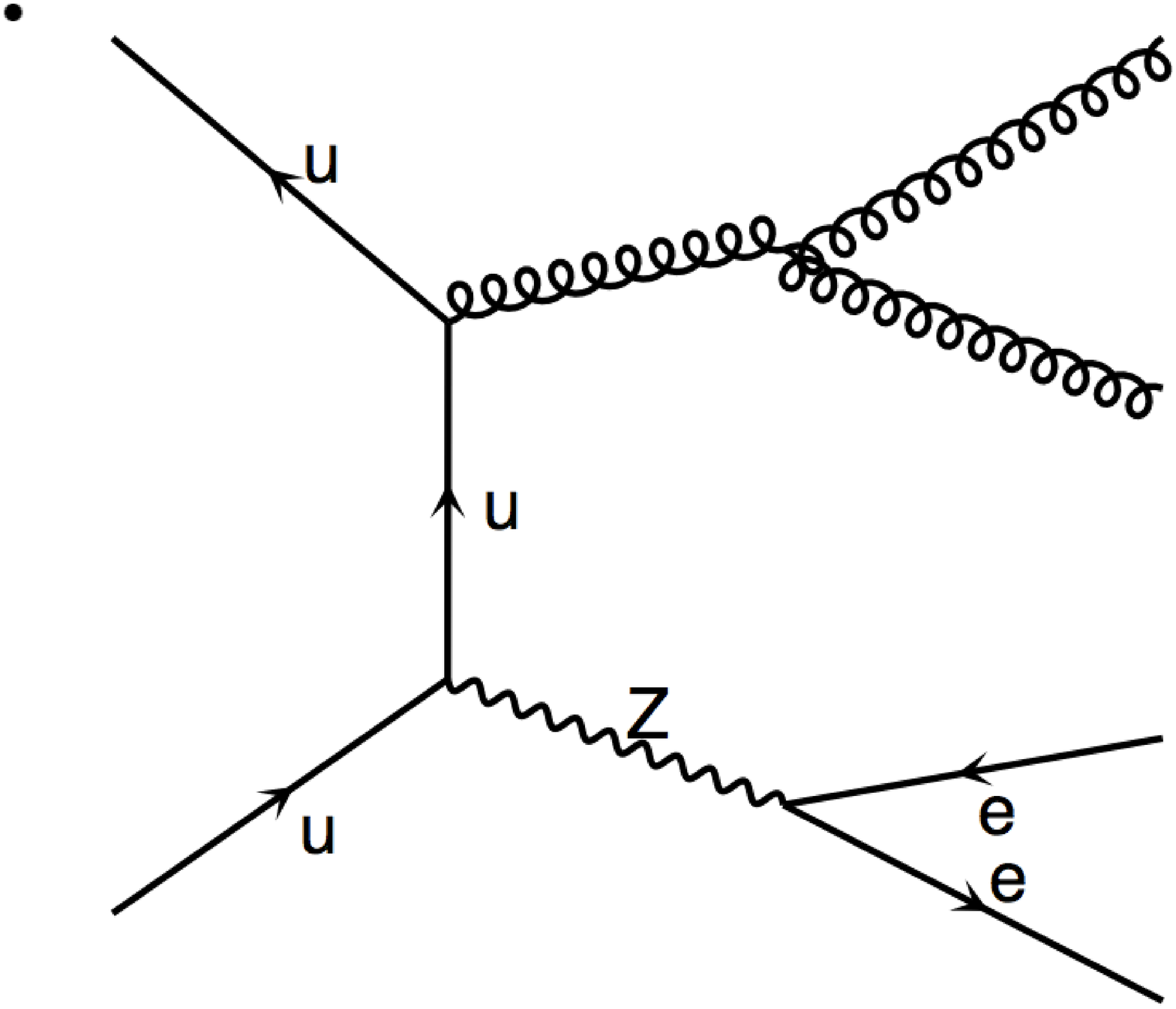} 
   \includegraphics[width = 0.3\textwidth]{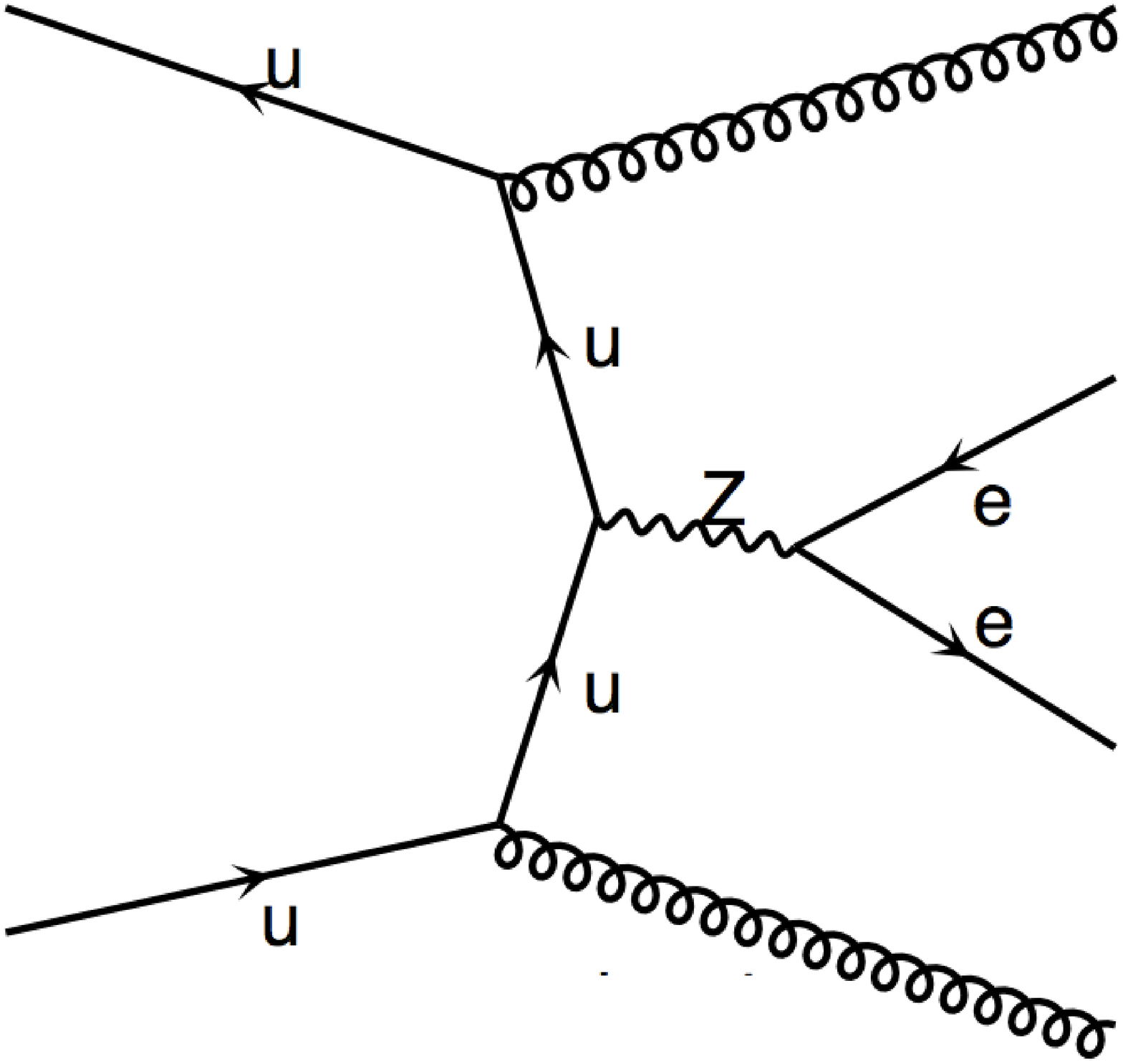} 
   \caption{Few  diagrams  contributing to the four fermions final state
                  at the $\alpha_{EW}^2 \times \alpha_s ^2$ leading order.}
   \label{fig:zjet}
\end{figure}

Other processes involving the production of a Z boson accompanied by jets
at the $\alpha_{EW}^2 \times \alpha_{s}^n$ order, with $n \ge 2$ (the so called Z+jets events),
have been considered as backgrounds.
An example of two, between the many Z production diagrams, are shown in figure \ref{fig:zjet}.
The cross section of this process is $\sigma= 2679$~pb, 
about 700 times larger than the electroweak one.

As a consequence, 
the selection cuts to suppress the backgrounds with respect to the VBF signal 
have been tightened with respect to the ones of table~\ref{tab:VBFRECOcuts} 
and selection cuts on the lepton transverse momentum, 
and on the opening angle between jets and between leptons in the transverse plane ($\Delta\phi$) 
have been added.
These selection cuts are listed in table \ref{tab:zjet}.

\begin{figure}[htbp]
   \centering
   \includegraphics[width = 0.45\textwidth]{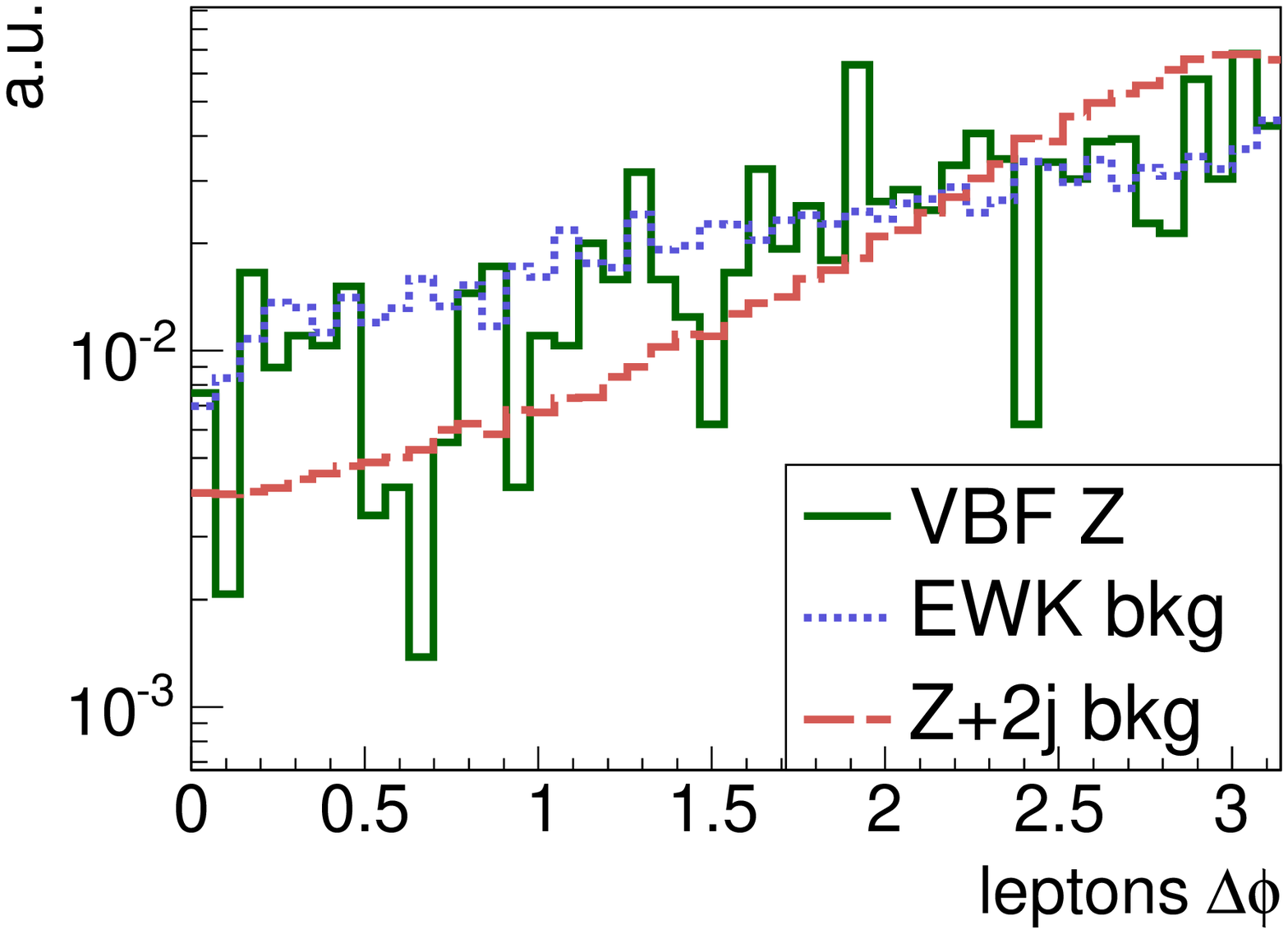} 
   \includegraphics[width = 0.45\textwidth]{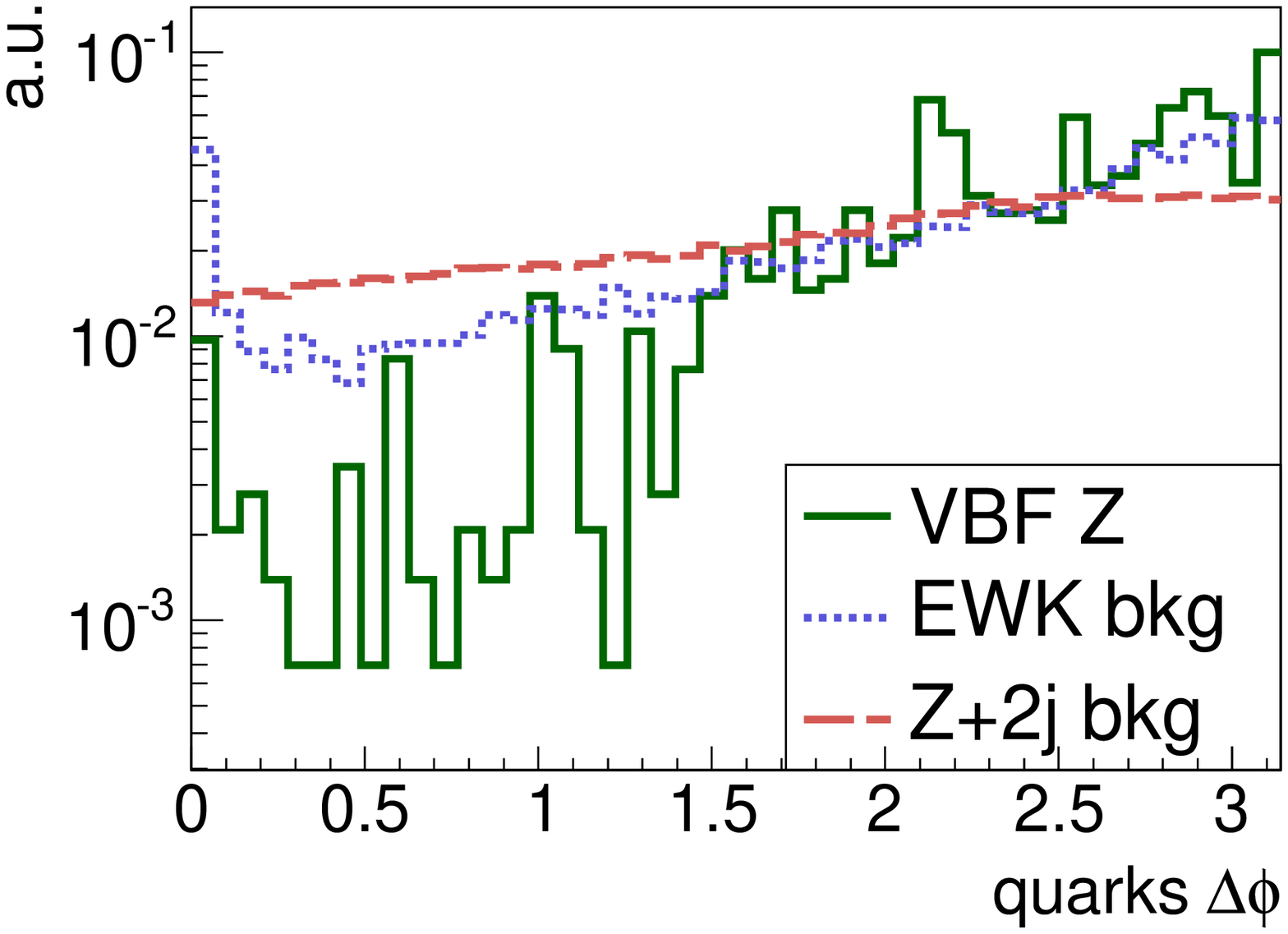} 
   \includegraphics[width = 0.45\textwidth]{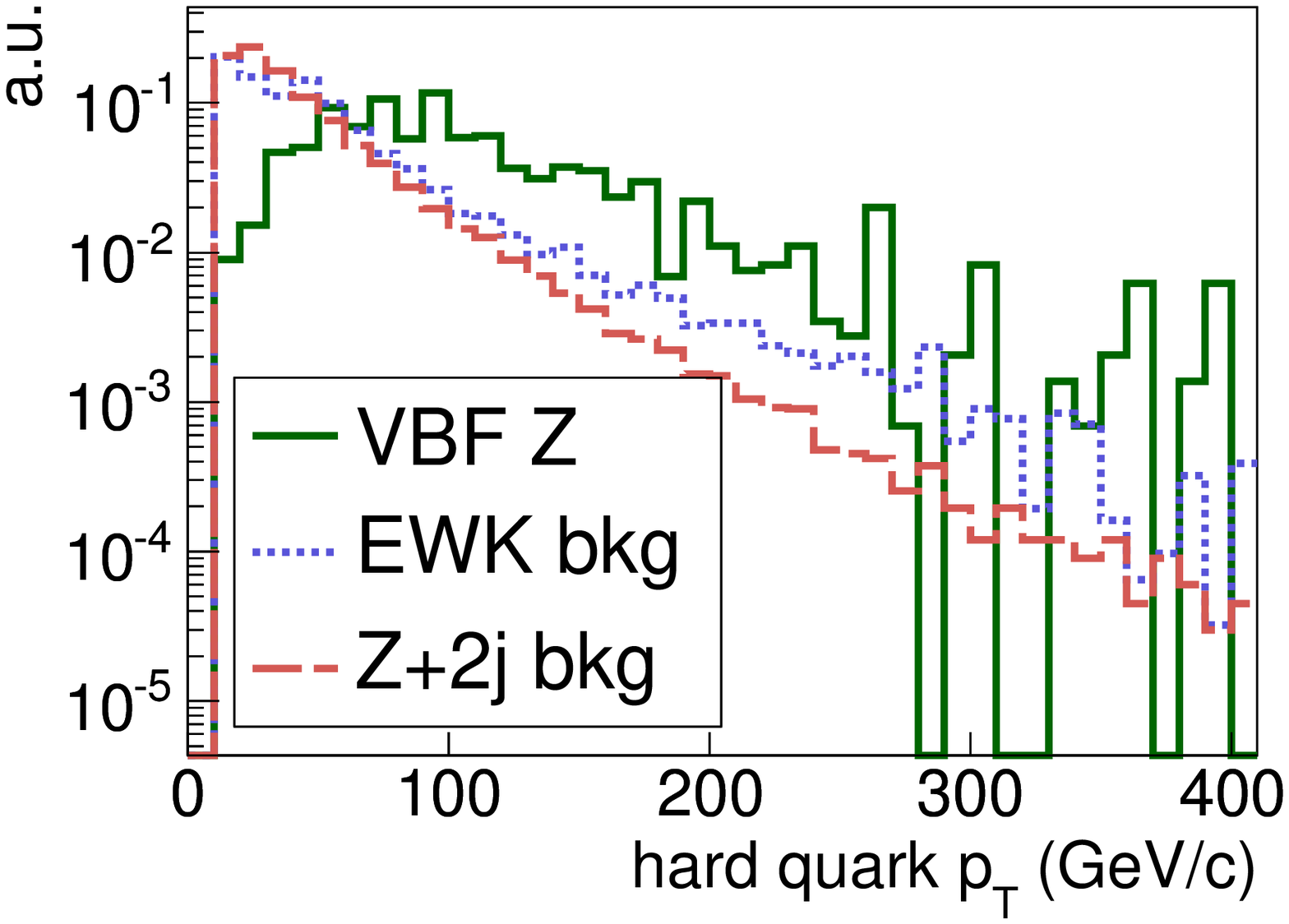} 
   \includegraphics[width = 0.45\textwidth]{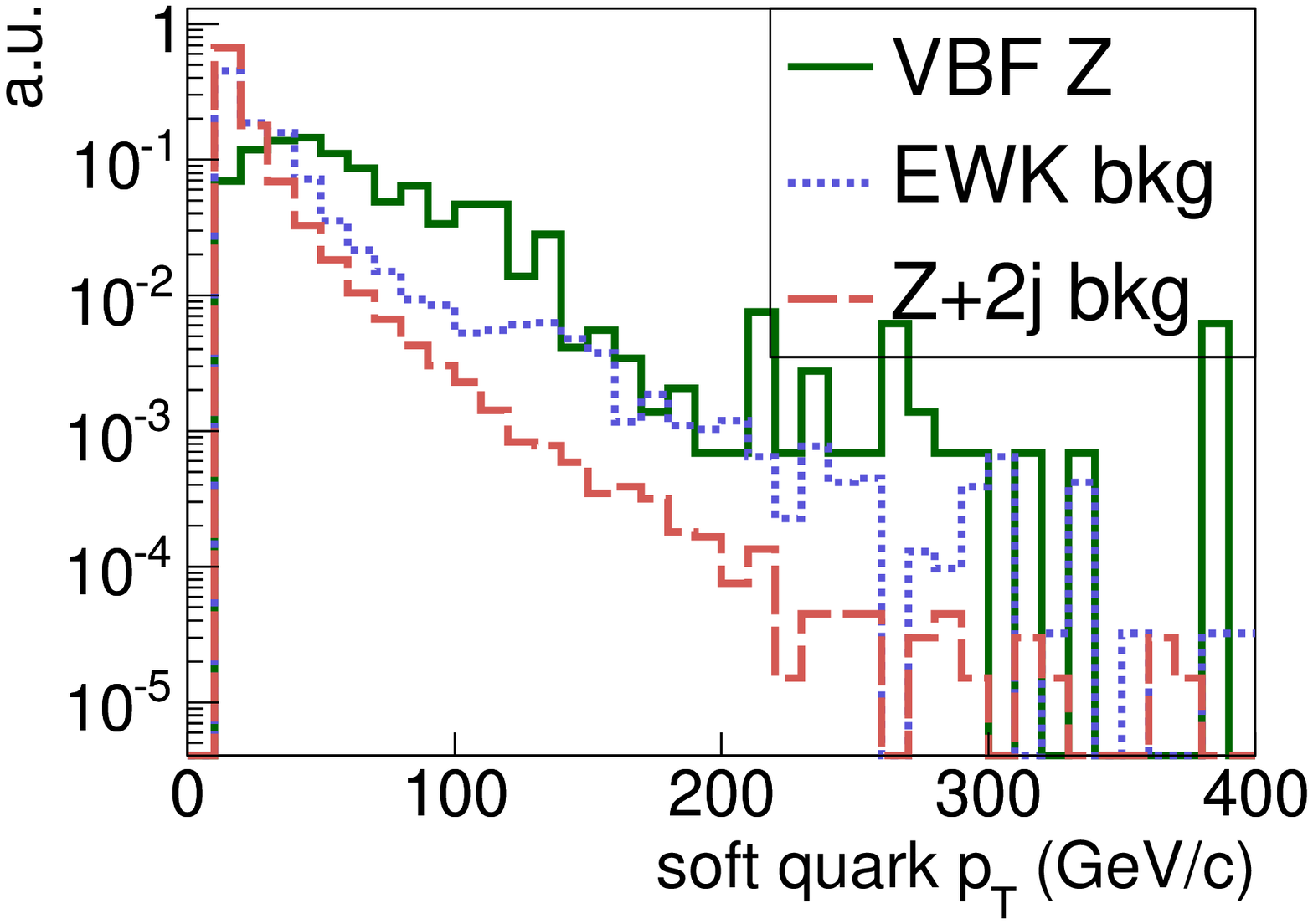} 
   \includegraphics[width = 0.45\textwidth]{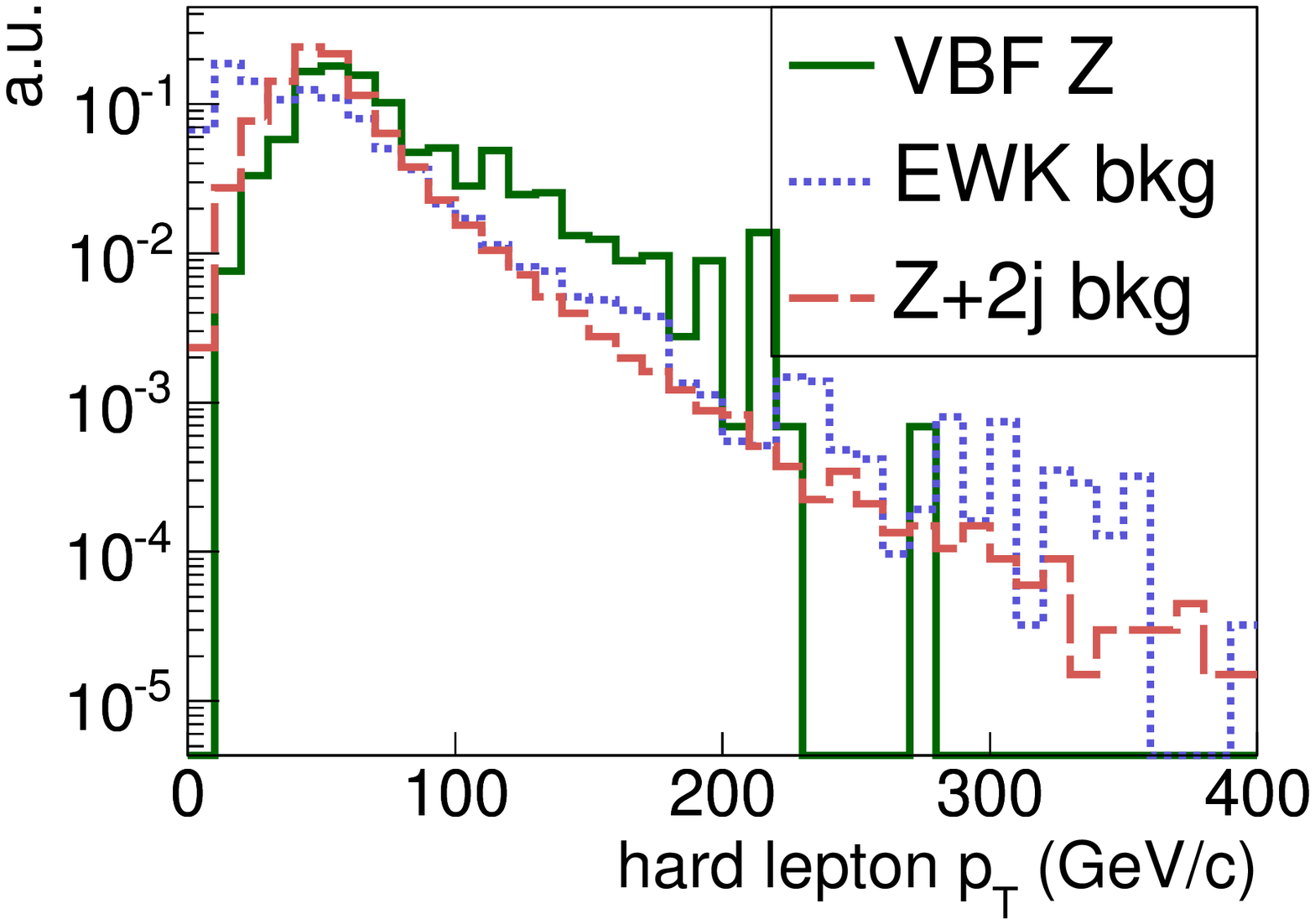} 
   \includegraphics[width = 0.45\textwidth]{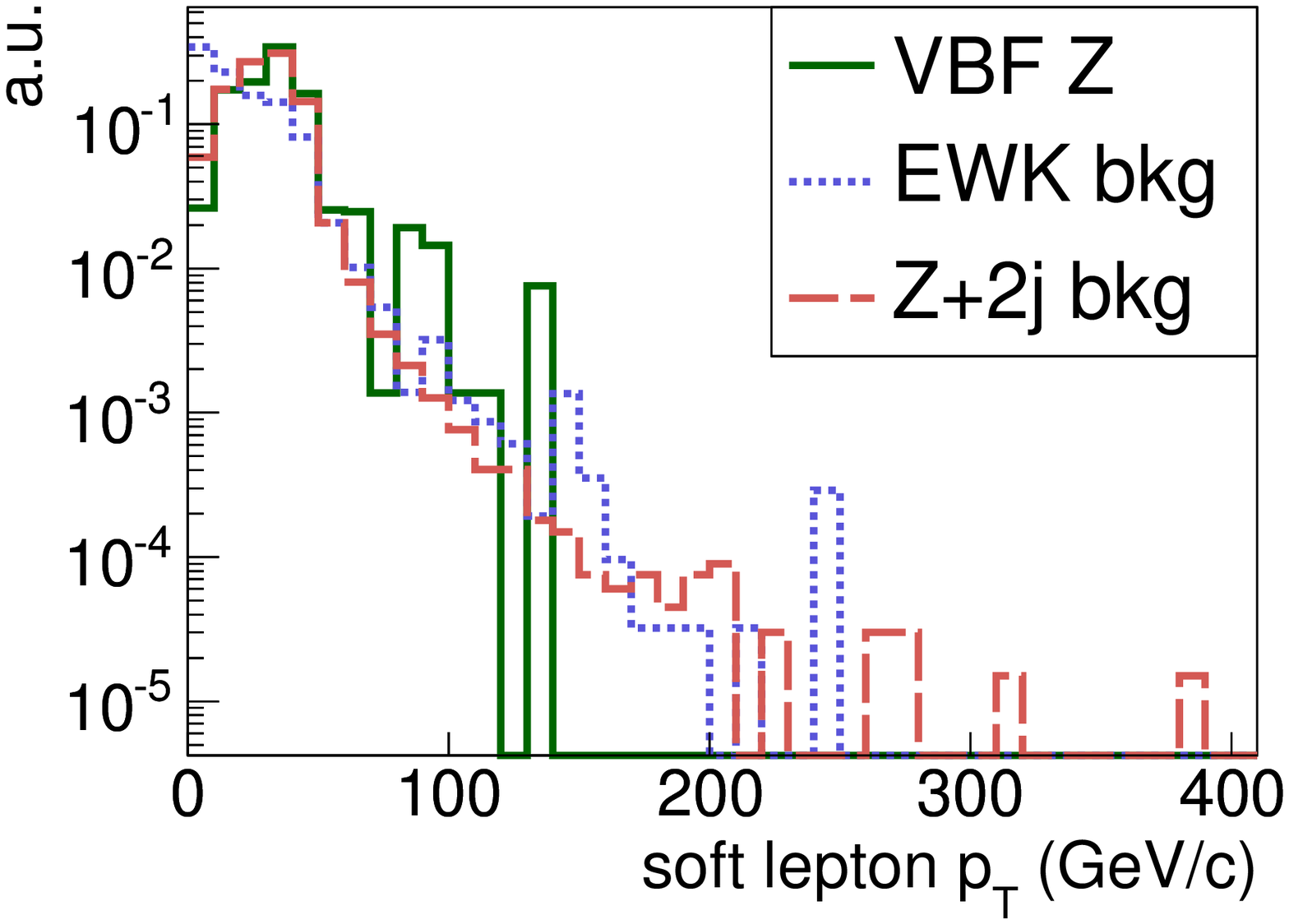} 
   \caption{
   The angular distance ($\Delta\phi$) between leptons and jets in the transverse plane is shown, 
   as well as the transverse momentum of the softest and hardest jet and lepton for the VBF signal (continuous green), 
   the pure electroweak backgrounds (short blue dashes), 
   and the Z+jets background (long red dashes), with arbitrary normalization.
   The last plot shows the two leptons invariant mass, for signal and backgrounds, normalized to their cross-sections.}
   \label{fig:sig-back-qcd}
\end{figure}

\begin{figure}[htbp]
   \centering
   \includegraphics[width = 0.65\textwidth]{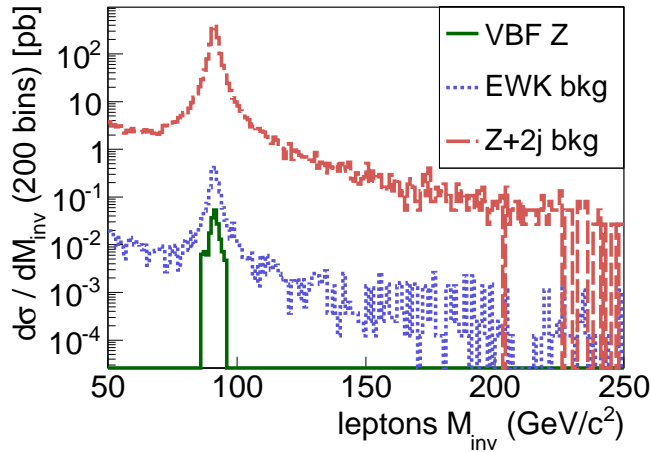} 
   \caption{
   The two final state leptons invariant mass, for the VBF signal (continuous green), 
   the pure electroweak backgrounds (short blue dashes), 
   and the Z+jets background (long red dashes), normalized to their cross-sections.}
   \label{fig:sig-back-qcd2}
\end{figure}

In figure \ref{fig:sig-back-qcd} the distribution of $\Delta\phi$ (for leptons and jets), 
the transverse momentum of the softest and hardest jet and lepton
are shown to justify the applied selection cuts.
In figure \ref{fig:sig-back-qcd2} the invariant mass of the two leptons is shown as well.

\begin{table}[htbp]
   \centering
   \begin{tabular}{|c|c|c|} 
\hline
selection var     & min     & max \\
\hline
$\Delta\eta_{jj}$ & 1.5     & -- \\ 
soft jet $p_T$    & 30 GeV  & -- \\
hard jet $p_T$    & 60 GeV  & -- \\
soft lepton $p_T$ & 15 GeV  & -- \\
hard lepton $p_T$ & 30 GeV  & -- \\
$\eta_j^*$        & 1       & 3 \\
$M_{ll}$          & 86 GeV  & 96 GeV \\
$M_{jj}$          & 400 GeV &  -- \\
$\Delta\phi_{jj}$ & 1.5     & 3.14 \\ 
$\Delta\phi_{ll}$ & 0       & 2.8 \\ 
$p_T(Z)$          & 30 GeV  & -- \\
\hline
   \end{tabular}
   \caption{ Selection cuts to suppress the Z+jets background.} 
   \label{tab:zjet}
\end{table}

\section{Results}

\begin{figure}[htbp]
   \centering
   \includegraphics[width = 0.65\textwidth]{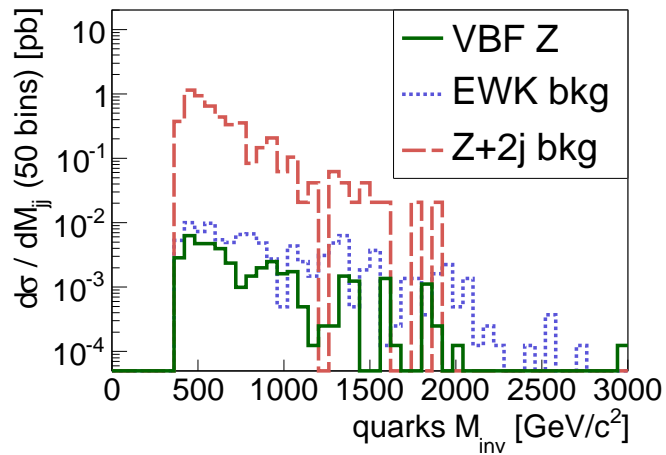} 
   \caption{
   Tag jets invariant mass distribution for the VBF signal (continuous green), 
   the EWK background (short blue dashes), 
   and the Z+jets background (long red dashes),
   after the selection cuts listed in table \ref{tab:zjet}. As can be seen, the Z+jets background 
   overwhelms the other samples.}
   \label{fig:final}
\end{figure}

The invariant mass of the two tag jets is shown in figure \ref{fig:final} after the final selection cuts.
Efficiencies and effective cross sections are summarized in table \ref{tab:VBFRECOHardCutsResults}.
The selected VBF signal results in a cross section of 0.04~pb, corresponding to 23\% efficiency.
The Z+jets background amounts to 5.14~pb of cross section, corresponding to 0.25\% of efficiency,
and is still two orders of magnitude larger than the signal. 
The signal over background ratio is 0.008.
No other obvious selections criteria have been found 
to enhance the signal over the dominant background at parton level.
Therefore, it is not possible to isolate the signal events from the dominant background
coming from single Z produced in association with jets coming from QCD.
\newline{}
After the hadronization,
the Z+jets background could be reduced by vetoing jet activity in the central part of the detector.
To perform a quick test,
after the parton showering and hadronization, jets have been reconstructed with an iterative cone algorithm.
The central jet veto reduces the Z+jets background by roughly 10\%,
not producing any significant improvement in the analysis.

\begin{table}[htbp]
   \centering
   \begin{tabular}{|l|c|c|c|} 
\hline
           & x-section          & effective x-section & efficiency \\
           &                    & after selections    &  \\
\hline
VBF signal & 0.18 pb            & 0.04 pb             & 0.23 \\
EWK bkg    & 3.85 pb            & 0.10 pb             & 0.027 \\
Z+jets bkg & 2679 pb            & 5.14 pb             & 0.0025 \\
\hline
S/B        & 6.71$\cdot10^{-5}$ & 0.008               & -- \\
\hline
   \end{tabular}
   \caption{VBF signal, EWK and Z+jets background cross sections 
   before and after the selection cuts  
   listed in table \ref{tab:zjet}.
   The signal and the electroweak background are defined by the MC selection cuts listed 
   in table \ref{tab:VBFMCcuts}.
   The final efficiencies and the signal over background (S/B) ratios are listed as well.}
   \label{tab:VBFRECOHardCutsResults}
\end{table}

\section{Conclusions}

One of the most powerful discovery channel for the Higgs Boson in all
the available mass range, is the one where the Higgs is produced via
Vector Boson Fusion.
The presence of the forward-backward jets and the absence of jets
activity in the central region are clear identification marks.
The possibility to identify single Z boson produced in VBF
 has been studied as a ``data driven'' method to measure the
efficiency to reconstruct and identify the  forward-backward jets and
the efficiency of the central jet veto.
Unfortunately it is not possible to isolate the signal events from the dominant background
coming from single Z produced in association with jets coming from QCD.





\begin{thebibliography}{9}

\bibitem{balle} A. Ballestrero et al. ``A complete parton level analysis of boson-boson scattering and electroweak symmetry breaking in $\ell\nu$ + four jets production at the LHC'' JHEP05(2009)015
\bibitem{ATLASTDR} ATLAS Collaboration, ``Expected Performance of the ATLAS Experiment: Detector, Trigger and Physics'', CERN-OPEN-2008-020 
\bibitem{CMSTDR} CMS Collaboration, ``CMS Physics, Technical design report'', CERN-LHCC-2006-001. 
\bibitem{dgreen} 
  D.~Green,
  ``Using q q Z events to 'calibrate' vector boson fusion at the LHC,''
  arXiv:hep-ex/0502009.
\bibitem{mad} 
J. Alwall \textit{et al.}, 
``MadGraph/MadEvent v4: The New Web Generation'', 
JHEP0709-028-2007, arXiv0706.2334
\bibitem{zeppenfeld} D. Zeppenfeld \textit{et al.}, Phys. Rev. D54:6680-6689,1996.
\end{thebibliography}
\end{document}